\documentclass[%
reprint,
 amsmath,amssymb,
 aps,
prl
]{revtex4-1}

\usepackage{amsmath}

\usepackage{graphicx}
\usepackage{dcolumn}
\usepackage{bm}

\begin{document}

\preprint{APS/123-QED}

\title{Possible Topological Phase Transition in Fe-Vacancy-Ordered $\beta$-Fe$_{4+\delta}$Se$_{5}$ Nanowires}

\author{Keng-Yu~Yeh$^{1,3,4}$}%
 \thanks{K.-Y. and T.-S. are co-first authors}%
\author{Tung-Sheng~Lo$^{1,\ast}$}%
 \thanks{tslo@phys.sinica.edu.tw}
\author{Chung-Chieh~Chang$^{1}$}%
\author{Phillip~Wu$^{1}$}%
\author{Kuei-Shu~Chang-Liao$^{4}$}%
\author{Ming-Jye~Wang$^{1,2}$}%
\author{Maw-Kuen~Wu$^{1}$}%
 \thanks{mkwu@phys.sinica.edu.tw}

\affiliation{$^1$Institute of Physics, Academia Sinica, Taipei 115, Taiwan}%
\affiliation{$^2$Institute of Astronomy and Astrophysics, Academia Sinica, Taipei 115, Taiwan}%
\affiliation{$^3$Taiwan International Graduate Student Program, Academia Sinica, Taipei 115, Taiwan}%
\affiliation{$^4$Department of Engineering and System Science, National Tsing Hua University, Hsinchu 300, Taiwan}%

\date{\today}

\begin{abstract}
We studied the electrical transport on $\beta$-Fe$_{4+\delta}$Se$_{5}$ single-crystal nanowires, exhibiting $\sqrt{5}\times\sqrt{5}$ Fe-vacancy order and mixed valence of Fe. We observed a first-order metal-insulator transition of the transition temperature at $\sim$28~K at zero magnetic field. The dielectric relaxation reveals that the transition is related to an energy gap expansion of $\sim$12~meV, involving the charge-orbital ordering. At nearly 28~K, colossal positive magnetoresistance emerges, resulting from the magnetic-field dependent shift of the transition temperature. Through the transition, the magnetotransport behavior transits from two-dimension-like to one-dimension-like conduction. The transition temperature demonstrates anisotropy with the $c$-axis as the preferred orientation in magnetic fields, suggesting the spin-orbital coupling. Our findings demonstrate the novel magnetoresistive transition intimating a topological transition in the Fe-vacancy-ordered $\beta$-Fe$_{4+\delta}$Se$_{5}$ nanowires. The results provide valuable information to better understand the orbital nature and the emergence of superconductivity in FeSe-based materials. 
\end{abstract}

\maketitle

The discovery of FeAs-based~\cite{Kamihara2008} and FeSe-based~\cite{Hsu2008} superconductors created an exciting platform for better understanding the physics of high-temperature superconductivity. The observation of wide-range of transition temperature ($T_\text{c}$), with the highest confirmed Cooper-pair formation temperature up to 75~K in monolayer FeSe film~\cite{Peng2014} is indeed intriguing and provides a unique opportunity to gain more insight into the origin of high-temperature superconductivity. 

The multiple-orbital nature of FeSe, combined with spin and charge degrees of freedom, results in the observation of many interesting phenomena such as nemacity~\cite{Fernandes2014,Yu2018}, orbital-selective Mott transition~\cite{Yi2013,Yu2013,Herbrych2018}, and orbital ordering~\cite{Yi2017}. There are suggestions that the orbital fluctuation may provide a new channel for realizing superconductivity~\cite{Saito2010,Kontani2010}. Direct determination of the correlation between the orbital nature of the low-energy electronic states and the superconducting gap is crucial to understand the superconductivity mechanism of the Fe-based superconductors. 

It has been a debate, similar to the cuprate superconductors, whether there exists an antiferromagnetic Mott insulating parent phase for FeSe superconductors. Chen et~al. reported the existence of the $\beta$-Fe$_4$Se$_5$ phase with $\sqrt{5}\times\sqrt{5}$ Fe-vacancy order and argued this $\beta$-Fe$_4$Se$_5$ phase to be the parent phase of FeSe superconductors~\cite{Chen2014}. A theoretical simulation shows that the ground state of $\sqrt{5}\times\sqrt{5}$ Fe-vacancy ordered $\beta$-Fe$_4$Se$_5$ has a pair-checkerboard antiferromagnetic order, and the Fe 3$d$ orbitals govern the low-energy physics~\cite{Gao2017}. Analogously, recent studies demonstrated unambiguously that the K$_2$Fe$_4$Se$_5$, which exhibits a $\sqrt{5}\times\sqrt{5}$ Fe-vacancy order accompanying with an antiferromagnetic order~\cite{Bao2011,Zhao2012}, is the parent phase of the superconductor K$_2$Fe$_{4+x}$Se$_5$~\cite{Wang2015,Wang2019}. Detailed studies of the Fe vacancy in K$_2$Fe$_{4+x}$Se$_5$ reveal that its order/disorder is closely associated with superconductivity. 
 
Fe-deficient $\beta$-Fe$_4$Se$_5$ is expected to exhibit mixed-valence, which may show correlations among the orders of charge, spin, and orbital, resembling magnetite~\cite{Jeng2004,Huang2006}, which exhibits the Verwey transition around 120~K. The metal-insulator transition and the drop in magnetization are the signatures of the Verwey transition~\cite{Walz2002}. In crystal-like $\beta$-Fe$_4$Se$_5$ sheets, a metal-insulator transition and a drop in magnetic susceptibility have been observed at low temperature~\cite{Chen2014}. A more detailed investigation of the Fe-deficient $\beta$-Fe$_{4+\delta}$Se$_5$ will provide clear picture of the physics underlying and contribute to better understanding the superconductivity in FeSe. Nonetheless, in spite of the success of synthesizing FeSe mesoscale materials having various Fe-vacancy orders~\cite{Chang2012,Chang2014,Chen2014}, Fe-vacancy-ordered $\beta$-Fe$_4$Se$_5$ is yet lack of sufficient investigation.

In this work, we present the electrical transport of the Fe-deficient $\beta$-Fe$_{4+\delta}$Se$_5$ single-crystal nanowires with $\sqrt{5}\times\sqrt{5}$-ordered Fe vacancies and mixed valence of Fe. A first-order metal-insulator (MI) transition with an onset transition temperature at $T$~$\sim$~28~K at zero magnetic field is observed. We investigate the magnetotransport and the dielectric relaxation to understand the carrier transport character through the transition. Our study demonstrates the possible topological transition in $\sqrt{5}\times\sqrt{5}$-Fe-vacancy-ordered $\beta$-Fe$_{4+\delta}$Se$_5$.  

\begin{figure}[b]
\includegraphics{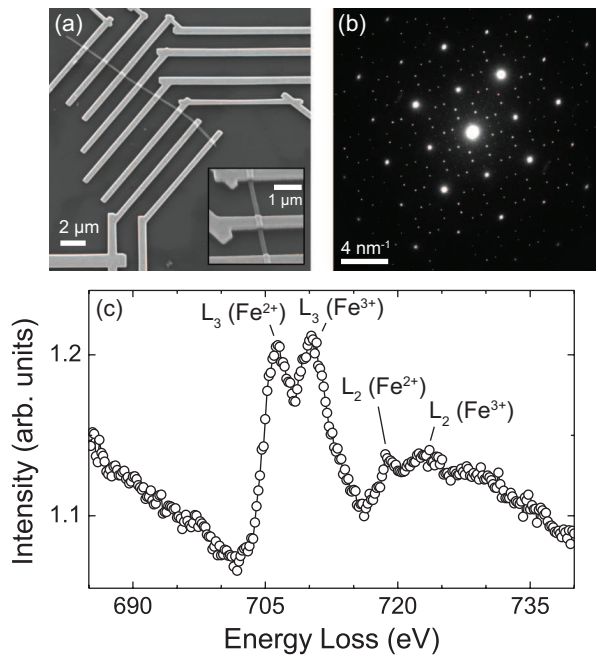}
\caption{(a) The SEM image of a Fe$_{4+\delta}$Se$_{5}$ nanowire after deposited electrical leads. The inset is the tilted SEM image, showing that the nanowire has width of $\sim$180~nm and thickness of $\sim$60~nm. (b) The SAED of a Fe$_{4+\delta}$Se$_{5}$ nanowire along the $c$-axis direction. (c) The EELS spectrum of Fe L$_{2,3}$-edges of a Fe$_{4+\delta}$Se$_{5}$ nanowire, showing the mixed valence of Fe. 
}
\end{figure}

The $\beta$-Fe$_{4+\delta}$Se$_5$ nanowires were grown by the on-film formation method~\cite{Chang2014}. The nanowires are typically rectangular strips with few-hundred-nanometer width and several-ten-micrometer in length as shown by the scanning electron microscope (SEM) image in Fig.~1(a). We used the 30-nm-thick Si$_3$N$_4$ membranes to hold the nanowires for the benefit of characterizing the nanowires \textit{in situ} by the transmission electron microscope (TEM). The electrical leads, patterned by electron beam lithography with 90~$^\circ\text{C}$ baked PMMA, were made of Au/Ti bilayer (150/30~nm) after the surface oxide layer of the nanowires removed by Ar plasma. The resistances of the nanowires were measured by the four-probe method and lock-in technique with an alternative-current~(AC) excitation less than 300~nA. Sample temperature and external magnetic field are controlled by PPMS (physical property measurement system), and the field orientation to the sample is changed by rotating the sample via a rotation insert.

For sample characterization, we used TEM to obtain the selected area electron diffraction (SAED) pattern, the scanning TEM energy dispersive X-ray spectrometer (STEM-EDS) for element analysis, and the electron energy loss spectroscopy (EELS) to determine the Fe valence. The element mappings show uniform distributions of Fe and Se on the nanowire as shown in Fig.~S1. The ratio between Fe and Se in this nanowire is 4.48/5 according to EDS analysis. As shown in Fig.~1(b),The SAED pattern reveals that the Fe$_{4+\delta}$Se$_5$ nanowire is a single crystal with a tetragonal structure, where the $b$-axis is parallel to the longitudinal axis of the nanowire, the $c$-axis is perpendicular and the $a$-axis is parallel to the wide side of the nanowire. The electron diffraction also clearly shows the super-structure related to the $\sqrt{5}\times\sqrt{5}$ Fe-vacancy order~\cite{Chen2014}. The $a$-axis lattice constant is estimated to be 3.73~{\AA}. The EELS data in Fig.~1(c) demonstrate the absorption peaks for the Fe L$_3$ edge (Fe$^{2+}$ at 706.4~eV and Fe$^{3+}$ at 710.2~eV) and the Fe L$_2$ edge (Fe$^{2+}$ at 718.4~eV and Fe$^{3+}$ at 723.5~eV), confirming the existence of mix-valence state of Fe$^{2+}$ and Fe$^{3+}$. 

\begin{figure}[b]
\includegraphics{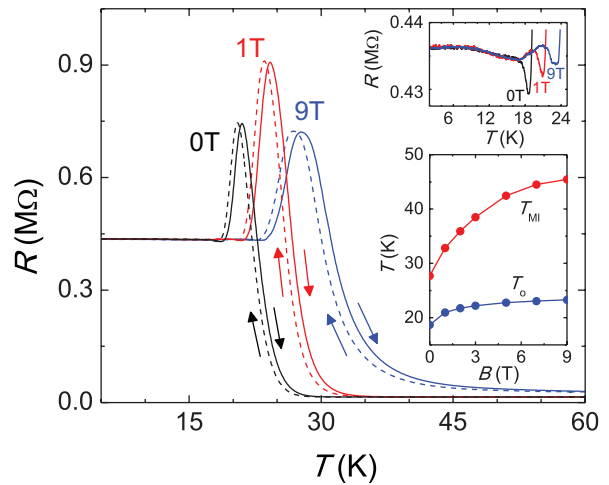}
\caption{$B$-field dependence of $R(T)$ with the field direction along the $c$-axis, including results measured in warming and cooling. The upper inset shows $R$ of warming at $T$~$<$~25~K. (b) $T$-$B$ phase diagram includes $T_\text{MI}$ and $T_\text{o}$ derived from $R(T)$ results. 
}
\end{figure}

We measured the temperature $(T)$ dependence of the resistance $(R)$ with the temperature control rate of 1~K/min and the AC excitation frequency at 37~Hz. Figure~2 shows the $R$ versus $T$ in cooling and warming measured at different magnetic fields $B$ parallel to the $c$-axis. The transition shows clear thermal hysteresis. At $B$~$=$~0~T, $R$ undergoes a MI transition with the transition onset temperature $T_\text{MI}$~$\sim$~28~K, where $T_\text{MI}$ is defined as the temperature at which the thermal hysteresis appears. As temperature decreases, $R$ drastically increases from $\sim$10~k$\Omega$ and reaches a resistance maximum $\sim$0.6~M$\Omega$ at $\sim$21~K, and then decreases to about 0.4~M$\Omega$ at $T_\text{o}$, which is defined as the temperature where the thermal hysteresis disappears. The thermal hysteresis of $R(T)$ indicates the transition is first-order, which is further confirmed by the applied magnetic field $B$, which enhances the thermal hysteresis loop. The upper inset in Fig.~2 shows more details of the $R(T)$ at $T$~$<$~24~K. A dip is found at $T$~$\sim$~19~K at $B$~$=$~0~T and shifts to higher temperature with increasing $B$ field. The resistance peak value does not change monotonically with $B$ fields. Notice also that the $R(T)$ plateau at $T$~$<$~17~K is solid under $B$ fields up to $B$~$=$~9~T, and reaches a fixed value of $\sim$0.436~M$\Omega$ at 2~K. The lower inset of Fig.~2 displays the phase diagram of the transition between the high-$T$ ($T$~$>$~$T_\text{MI}$) and low-$T$ ($T$~$<$~$T_\text{o}$) phases. Both $T_\text{MI}$ and $T_\text{o}$ increase with increasing $B$.

\begin{figure}[b]
\includegraphics{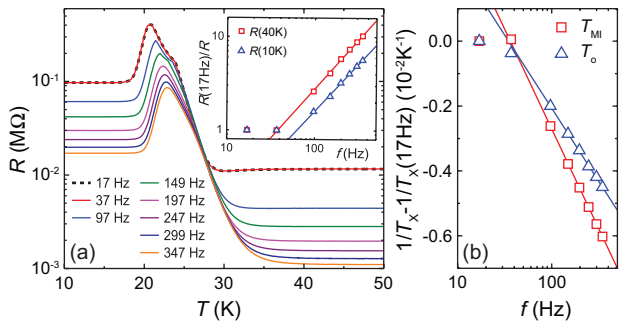}
\caption{The AC-excitation frequency, $f$, dependence on the first-order transition. (a) $R(T)$ curves measured in warming at different $f$. The curves of 17~Hz and 37~Hz are almost overlapped. The inset shows the $f$ dependence of conductance $1/R(10\text{K})$ and $1/R(40\text{K})$, normalized by the values at $f$~$=$~17~Hz. The solid lines show $1/R$~$\propto$~$f^{\alpha}$, where $\alpha$ is 1.06 for $1/R(40\text{K})$ and 1.02 for $1/R(10\text{K})$. (b) $1/T_\text{X}(f)$~$-$~$1/T_\text{X}(17\text{Hz})$ for $T_\text{X}$ is $T_\text{MI}$ or $T_\text{o}$. $T_\text{MI}(17\text{Hz})$~$=$~27.4~K and $T_\text{o}(17\text{Hz})$~$=$~17.5~K. The solid lines depict the Arrhenius relation $f(T_\text{X})$~$\propto$~exp$(-E_\text{a}/k_\text{B}T_\text{X})$, where $E_\text{a}$ is 32.3~meV for $T_\text{MI}$ and 44.0~meV for $T_\text{o}$. 
}
\end{figure}

To gain more insight into the nature of the transition, we investigated the AC-excitation frequency $(f)$ dependence of $R(T)$ at zero $B$ field. The data were measured from a different section of the same nanowire for the $B$-field dependent experiment. The $R(T)$ at different $f$ are shown in Fig.~3(a). $R$ decreases with increasing $f$ while being weakly $f$-dependent at $T$~$\sim$~28~K. As shown in the inset of Fig.~3(a), as $f$~$>$~37~Hz, $1/R$ shows power-law $f$ dependence, i.e. $1/R$~$\propto$~$f^{\alpha}$ with $\alpha$~$\sim$~1, for both $T$~$<$~$T_\text{o}$ and $T$~$>$~$T_\text{MI}$. The power-law spectrum is known as a fingerprint of dielectric relaxation for hopping systems~\cite{Jonscher1999,Zvyagin2006}. The distinct weak $f$ dependence at $T$~$\sim$~28~K, which is around $T_\text{MI}$ at zero field, suggesting that it is the critical point of the phase transition. The kinks occur in the transition regime at $f$~$\leq$~100~Hz may result from the discrete jumps in order parameters, which has also been observed in the MI transition of mesoscale strongly correlated systems~\cite{Zhai2006,Sharoni2008,Uhlir2016}. 

Figure~3(b) shows the $f$ dependence of $T_\text{MI}$ and $T_\text{o}$. When $f$~$>$~37~Hz, $T_\text{MI}$ and $T_\text{o}$ shift with the Arrhenius law $f(T_\text{X})$~$\propto$~exp$(-E_\text{a}/k_\text{B}T_\text{X})$, where $E_\text{a}$ is the activation energy and $T_\text{X}$ is $T_\text{MI}$ or $T_\text{o}$. The Arrhenius law indicates the shift is thermally activated. The extracted $E_\text{a}$'s are 32.3~meV and 44.0~meV for $T_\text{MI}$ and $T_\text{o}$, respectively. The $E_\text{a}$ enhancement across the transition with decreasing $T$ is estimated to be ${\Delta}E$~$=$~$E_\text{a}(T_\text{o})-E_\text{a}(T_\text{MI})$~$\sim$~12~meV. The ratio of $T_\text{MI}$ to ${\Delta}E$ is $\sim$30K/12meV~$=$~$\sim$2.5~K/meV, which is comparable to $\sim$125K/50meV~$=$~$\sim$2.5~K/meV, the ratio of $T_\text{V}$ to the energy gap change deduced from optical conductivity~\cite{Pimenov2005} and photoemission spectrum~\cite{Schrupp2005,Park1997} for single-crystal magnetite. In magnetite studies, the energy gap expansion refers to the increase of $E_\text{a}$ when lowering $T$ over the Verwey transition~\cite{Park1997,Pimenov2005,Schrupp2005,Ramos2008}, and the $E_\text{a}$ values extracted from different experimental approaches~\cite{Kuipers1976,Park1997,Ramos2008,Gooth2014}, including the $f$-dependent shift in AC conductance, are comparable. Thus, ${\Delta}E$ observed in the $\beta$-Fe$_{4+\delta}$Se$_5$ nanowire could be attributed to an energy gap expansion alike to the energy gap expansion observed in the Verwey transition, intimating that charge-orbital ordering may play an important role in the MI transition~\cite{Huang2006,Jeng2004}. 

\begin{figure}[b]
\includegraphics{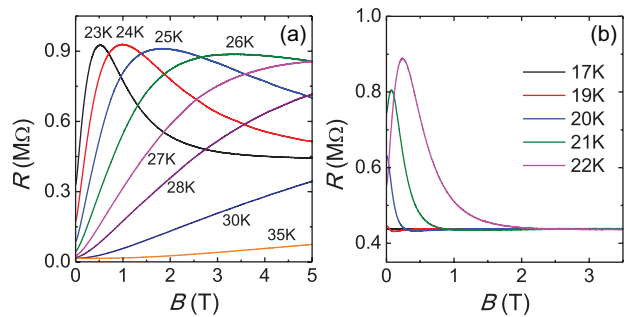}
\caption{$R$ versus $B$ with $B$ fields parallel to the $c$-axis at different $T$ from 23~K to 35~K (a) and from 17~K to 22~K (b). 
}
\end{figure}

As mentioned earlier, the MI transition can be substantially enhanced by applying magnetic fields, as shown in Fig.~2. Consequently, we decided to carry out detailed magnetotransport measurement. Figures~4(a) and 4(b) show the magnetoresistance $R(B)$ of the nanowire at different temperatures with the $B$-field direction parallel to the $c$-axis. At $T$ well above zero-field transition temperature $T_\text{MI}(0\text{T})$, i.e. $T$~$>$~30~K, $R(B)$ exhibits parabolic-like dependence. At $T$ above but close to $T_\text{MI}(0\text{T})$, e.g. $T$~$=$~30~K, $R(B)$ shows linear $B$-field dependence. As 20~K~$<$~$T$~$<$~26~K, a resistance peak in $R(B)$ emerges owing to the $B$-dependent shift of the $R(T)$ peak, and the position of the peak maximum decreases with lowering $T$. Colossal positive magnetoresistance with ${\Delta}R(B)/R(0\text{T})$~$>$~10, where ${\Delta}R(B)$~$=$~$R(B)-R(0\text{T})$, is observed in 26~K~$<$~$T$~$<$~34~K. The colossal magnetoresistance is due to the rise of $T_\text{MI}$ by applying $B$ fields. Strikingly, $R(B)$ at $T$~$\leq$~17~K is essentially independent on $B$. We do not find obvious magnetic hysteresis in magnetoresistance, being consistent with the picture of antiferromagnetism in $\sqrt{5}\times\sqrt{5}$ Fe-vacancy-ordered $\beta$-Fe$_4$Se$_5$~\cite{Chen2014,Fang2016,Gao2017}. 

\begin{figure}[b]
\includegraphics{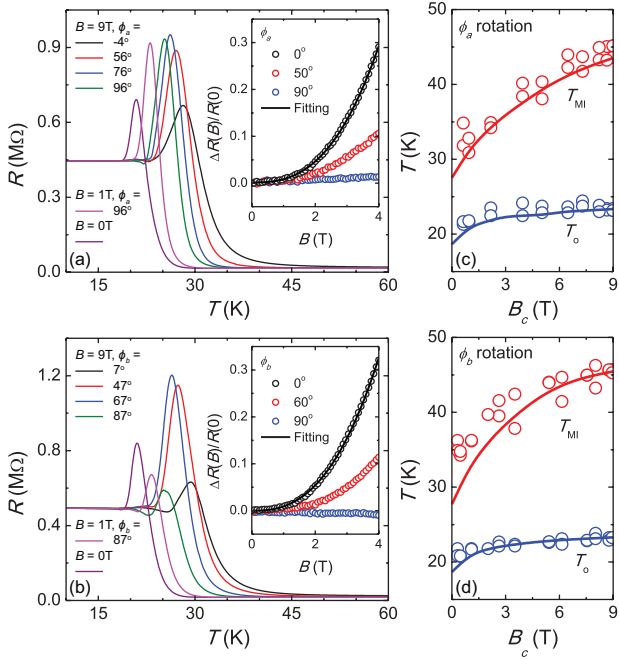}
\caption{$B$-field orientation dependence. (a) and (b) $R(T)$ measured in warming at $B$ fields of different $\phi_a$ and $\phi_b$, respectively.The insets show ${\Delta}R(B)/R(0\text{T})$ of $T$~$=$~40~K at different $\phi_a$ and $\phi_b$, respectively. (c) and (d) $T_\text{MI}$ and $T_\text{o}$ versus $B_{c}$, the $B$-field component in the direction along the $c$-axis, in $\phi_a$ and $\phi_b$ rotations, respectively. The solid lines represent $T_\text{MI}$ and $T_{o}$ deduced when the $B$ fields parallel to the $c$-axis. 
}
\end{figure}

We further investigated the $B$-field orientation dependence of the transition at $B$~$=$~9~T with the nanowire rotated by an angle $\phi_a$ ($\phi_b$) with the rotation axis along the $a$-axis ($b$-axis). The $B$-field direction is parallel to the $c$-axis as $\phi_a$ ($\phi_b$) is zero. The data of $\phi_a$ and $\phi_b$ dependences were acquired from two different sections of the same nanowire. Figures~5(a) and 5(b) show the $R(T)$ results at $B$~$=$~9~T at different $\phi_a$ and $\phi_b$, respectively. As the $B$-field direction changes from parallel to perpendicular to the $c$-axis, the positions of the MI transition and the resistance peak shift to lower $T$. Figures~5(c) and 5(d) show the $T_\text{MI}$ ($T_\text{o}$) versus $B_c$, the $B$-field components parallel to the $c$-axis, where $B_c$~$=$~$9\text{T}\times|\text{cos}(\phi_a)|$ and $B_c$~$=$~$9\text{T}\times|\text{cos}(\phi_a)|$ for different $\phi_a$ and $\phi_b$ at $B$~$=$~9~T, respectively. The $T_\text{MI}$ ($T_\text{o}$) values of different $\phi_a$ and $\phi_b$ scatter along the trace of $T_\text{MI}$ ($T_\text{o}$) versus $B_c$~$=$~$B$, where the $B$-field direction is parallel to the $c$-axis. In other words, $T_\text{MI}$ ($T_\text{o}$) is enhanced anisotropically in $B$ fields with the preferred orientation along the $c$-axis. This observation indicates that the transition may be associated with the spin-orbital coupling in the Fe-vacancy-ordered $\beta$-Fe$_{4+\delta}$Se$_5$ . 

The insets in Fig.~5(a) and 5(b) show ${\Delta}R(B)/R(0\text{T})$ at $T$~$=$~40~K of different $B$-field orientations as an example to demonstrate the magnetotransport in the high-$T$ phase. When $B$ fields orient along the $c$-axis, ${\Delta}R(B)/R(0\text{T})$~$\propto$~$B^{\gamma}$, where $\gamma$~$=$~2.7~($\gamma$~$=$~2.6) in the inset of Fig.~5(a)~(5(b)), which is not a conventional Lorentzian magnetoresistance. ${\Delta}R(B)/R(0\text{T})$ decreases when $B$ fields orient from parallel to perpendicular to the $c$-axis. When $B$ fields are perpendicular to the $c$-axis, $\left|{\Delta}R(B)/R(0\text{T})\right|$~$<$~0.01 and is weakly $B$-dependent, showing a two-dimensional-like (2D-like) magnetotransport character and indicating the stripe-shaped nanowire as a quasi-2D system. At $T$~$<$~17~K, $R$ values under applying $B$ fields is rigid regardless of changing $B$-field orientation and magnitude, behaving like one-dimensional (1D) conduction of fully spin polarization~\cite{Chen2008,Chen2012}, reminiscent of a Wigner-like phase~\cite{Wu2012,Zhang2013}. The $B$-field orientation dependence of $R$ peak value is complicated, as shown in Fig.~5(a) and 5(b). When $B$ fields are parallel to the $a$-axis ($b$-axis), the peak height increases (decreases) with increasing $B$. The $R$ peak emerging in the transition is in the phase coexistence regime of the first-order transition. The peak-like feature is also observed in three-dimensional (3D) topological insulators and is regarded as a sign of the existence of the topological surface state~\cite{Cai2018,Wolgast2013,Luo2015,Dankert2018}. The high-$T$ 2D-like phase and the low-$T$ 1D-like phase may be different in topology. $R$ peak feature of the nanowire at $T$~$<$~$T_\text{MI}$ may reveal the emergence of the topological 1D channels. The magnetoresistive transition suggests a possible topological phase transition. 

We studied the electrical transport of tetragonal $\beta$-Fe$_{4+\delta}$Se$_5$ nanowires possessing $\sqrt{5}\times\sqrt{5}$ Fe-vacancy ordering and mixed-valance state of Fe$^{2+}$ and Fe$^{3+}$. The first-order MI transition of $T_\text{MI}$~$\sim$~28~K observed at zero $B$ field, and a Verwey-like energy gap expansion of $\sim$12~meV, suggesting that charge-orbital ordering may involve the transition, is extracted from the thermal activation of the dielectric relaxation. When applying external $B$ fields, the MI transition shifts toward higher temperatures remarkably, resulting in the colossal positive magnetoresistance, and the thermal hysteresis gets more pronounced. The $B$-dependent rise of $T_\text{MI}$ and $T_\text{o}$ shows anisotropy with the preferred orientation parallel to the $c$-axis, indicating that the spin-orbital coupling in the nanowire is significant. Based on the results of magnetoresistance, the high-$T$ phase shows 2D-like conduction, and the low-$T$ phase shows the rigidity under external $B$ fields, behaving like fully spin-polarized 1D conduction. The high-$T$ and low-$T$ phases might be topologically distinct, demonstrating the possible topological phase transition in $\sqrt{5}\times\sqrt{5}$ Fe-vacancy-ordered $\beta$-Fe$_{4+\delta}$Se$_5$ nanowires. Our results could provide critical information for better understanding the origin of superconductivity in FeSe-based superconductors and the orbital-related physics in the FeSe-based materials. 

\begin{acknowledgments}
We thank Dr.~Peramaiyan~Ganesan and Dr.~Yen-Fu~Liu of the Institute of Physics, Academia Sinica, for valuable discussion. The work is supported by the Ministry of Science and Technology under Grant No.~MOST108-2633-M-001-001 and Academia Sinica Thematic Research Grant No.~AS-TP-106-M01.
\end{acknowledgments}


%

\end{document}